\documentclass{ws-mpla}
\usepackage[super,compress]{cite}
\usepackage{graphicx}
\usepackage[breaklinks]{hyperref}
\hypersetup{colorlinks,urlcolor=black,citecolor=black,linkcolor=black,filecolor=black}
\usepackage{color}
\begin{document}

\markboth{S. Ghosh }
{GWs from bosonic clouds of rotating BHs}

\catchline{}{}{}{}{}

\title{Gravitational waves from superradiant instabilities of rotating black holes
}

\author{\footnotesize Shrobana Ghosh}

\address{Gravity Exploration Institute, \\Cardiff University, CF10 3AT,\\
United Kingdom\\
ghoshs9@cardiff.ac.uk}

\maketitle

\pub{Received (Day Month Year)}{Revised (Day Month Year)}

\begin{abstract}
Direct detection of gravitational waves from several compact binary coalescences has ushered in a new era of astronomy. It has opened up the possibility of detecting ultralight bosons, predicted by extensions of the Standard Model, from their gravitational signatures. This is of particular interest as some of these hypothetical particles could be components of dark matter that are expected to interact very weakly with Standard Model particles, if at all, but they would gravitate as usual. Ultralight bosons can trigger superradiant instabilities of rotating black holes and form bosonic clouds that would emit gravitational waves. In this article we present an overview of such instabilities as gravitational wave sources and assess the ability of current and future detectors to shed light on potential dark matter candidates.  

\end{abstract}

\ccode{PACS Nos.: include PACS Nos.}

\section{Introduction}	
The most remarkable thing about Newton's theory of gravitation was, perhaps, his insight that the force responsible for objects falling on the ground is the same as the force responsible for motion of heavenly bodies. It revolutionized the field of astronomy and led to remarkable discoveries for centuries to come. A little over two centuries later, Einstein made an observation of equal remark whereby he proposed that gravity is not a force, it is merely geometry of spacetime. In addition to the impact of this new theory on conventional astronomy through bending of light by gravity, its other remarkable contribution to the field was realized in the form of a new kind of astronomy -- gravitational wave (GW) astronomy. Instead of using light as the messenger, the GW ``telescopes" use perturbations of spacetime to receive messages from distant astrophysical sources. 

One of the most fascinating consequence of Einstein's theory of gravity is the prediction of black holes (BHs), regions of spacetime where curvature diverges (singularities) hidden behind a surface from which not even light can escape (event horizon). These form, arguably, the most interesting astrophysical objects that arise as solutions to Einstein's equations for a vacuum spacetime. The first exact solution of Einstein's theory was discovered by Schwarzschild in 1916 ~\cite{1916SPAW.......189S} for the spacetime around a static, spherically symmetric object and this was identified as the metric for a non-rotating BH. Nearly half a century later, Kerr ~\cite{PhysRevLett.11.237} discovered the solution for the spacetime around a rotating BH and this is by far the most astrophysically relevant one.

The line element for a Kerr BH in the Boyer-Lindquist coordinates can be written as,
\begin{equation}\label{Kerrmetric}
\begin{split}
ds^2= - \Bigg(1-\frac{2Mr}{\Sigma}\Bigg)dt^2 - \frac{4Mar \textrm{sin}^2\theta}{\Sigma}dt d\phi + \frac{\Sigma}{\Delta}dr^2 +\Sigma d\theta^2\\
 + \textrm{sin}^2\theta\Bigg(r^2+a^2+\frac{2Mr}{\Sigma}a^2 \textrm{sin}^2\theta\Bigg)d\phi^2, 
\end{split}
\end{equation}
where $\Delta=r^2-2Mr+a^2, \,\, \Sigma=r^2+a^2 \textrm{cos}^2\theta$,\,\, \textit{M} is the mass of the BH, and $a=J/M$ is the Kerr angular momentum parameter, related to the  dimensionless spin parameter by $\chi=a/M$ (we use units where $G=c=1$). It is beneficial to write the metric in these coordinates as it immediately becomes apparent that the metric is not diagonal (i.e., the $dt d\phi$ term), which attributes some interesting features to these BHs. 
The event horizon of a Kerr BH is located at, 
\begin{equation}\label{horizon}
r_+=M+\sqrt{M^2 - a^2}.
\end{equation}

The angular velocity of the horizon can be shown to be,

\begin{equation}\label{horizonvelocity}
\Omega_H = \frac{a}{2Mr_+}
\end{equation}
An observer on a rotating spacetime will always have a non-zero angular velocity due to an effect called frame dragging, so the inertial frames nearest to the event horizon are rotationally dragged by it at nearly $\Omega_H$. To appear stationary with respect to infinity an observer would have to move in a way to counter the ``drag". The velocity with which such an observer needs to travel increases as one gets closer to the horizon, peaking at the static limit where the requisite speed is that of light. The region between the event horizon and the static limit is known as the ergosphere or ergoregion. Mathematically, it is the surface at the which the $g_{tt}$ component of the metric vanishes i.e.,
\begin{equation}\label{staticlimit}
r_e (\theta) = M + \sqrt{M^2-a^2\cos^2\theta};
\end{equation}
due to the $\theta$ dependence, this has the shape of an oblate spheroid (cf. Fig. 5 on~\refcite{Brito:2015oca}).

An interesting feature of the ergosphere is the existence of negative energy states within it. This can be understood by looking at the behavior of the $g_{tt}$ component of the metric across the static limit (see~\refcite{2003gieg.book.....H} for detailed explanation). The independence of the Kerr metric components on both $t$ and $\phi$ implies there are two Killing vectors for this spacetime -- a timelike Killing vector that corresponds to conservation of energy, and a spacelike Killing vector, corresponding to conservation of angular momentum. Outside the static limit the $g_{tt}$ component is negative (as per the convention chosen here, $-+++$), and it switches sign on crossing the static limit. The norm of the timelike Killing vector is no longer negative and it is therefore spacelike within the ergoregion. The conserved energy can now essentially masquerade as a momentum component and assume negative values leading to the existence of negative energy states.

This prompted Penrose to propose a thought experiment with fascinating consequences \cite{Penrose:1969pc}~. He hypothesised that a particle falling into Kerr BH may divide into two parts in the ergoregion. One of these parts can follow a negative energy orbit and cross the event horizon, while the other part emerges with a higher energy than that of the original particle. It has been shown that the energy of the outgoing particle for favorable conditions (eg., charged particles in an ambient electromagnetic field \cite{1985ApJ...290...12W, 1986ApJ...301.1018W, 1986ApJ...307...38P}, collisional Penrose process \cite{1975ApJ...196L.107P , PhysRevLett.103.111102}) can be significantly amplified. It is important to note here, that this is not in violation of conservation of energy as the particle gains energy at the expense of the BH losing some; for a Kerr BH that amounts to the BH spinning down.

Zel'dovich reimagined this from a field theory perspective \cite{zeldovich1} and proposed that an ingoing wave can suffer a similar fate within the ergoregion, leading to a transmitted part that falls into the BH while the reflected part gets amplified. This is known as \textit{superradiance} and it can occur for any dissipative system, which allows for partial absorption of the ingoing energy. This means any BH with an ergoregion can exhibit superradiance. As rotating BHs are considered to be the most astrophysically relevant ones, for the purposes of this review we will focus only on Kerr BHs (see \refcite{Brito:2015oca} for a detailed review on BH superradiance). 

While BH superradiance is yet to be confirmed by observation, this phenomenon has already been observed in a system \textit{analogous} to rotating BHs. Analogue BH models were first proposed by Unruh \cite{Unruh:1981}~. He noted that for an irrotational flow, a fluid may have surfaces where the speed of the flow exceeds the speed of sound. This surface can then behave like a BH horizon for sound waves and exhibit superradiance for the appropriate conditions. Indeed this was observed for plane water waves in a bathtub vortex \cite {Torres:2016iee}~. This is perhaps the best effort that can be made to convince ourselves of \textit{direct} observation of BH superradiance. Fortunately, the scope of an indirect observation is much more varied and promising.

Press and Teukolsky noted that if a \textit{mirror} was placed outside a BH then the outgoing wave could get reflected back towards the BH undergoing multiple amplification leading up to an instability, called a \textit{BH bomb}\cite{Press:1972zz}~. The effect of such mirroring boundary conditions can be produced by assigning a mass to the field. The superradiant instability manifests itself in the form of a cloud around the BH, rotating with it. Together with the BH, the cloud forms a “gravitational atom" with the energy spectrum being quantized much like a hydrogen atom \cite{Detweiler:1980uk, Dolan}~. Note that the quantum-mechanical analogy used here is mainly for the convenience of labelling the energy levels with numbers analogous to quantum numbers --  $n, \ell$ and $m$; these levels would be occupied by large number of particles allowing us to treat the systems purely classically. The particle analogy of the repeated amplification can be realised as an energy level being populated at the expense of the BH losing its rotational energy. Apart from borrowing the concept of quantized energy levels, we will stick to a purely classical picture (albeit relativistic) for the purposes of this review. 

It was first noted by Arvanitaki et.al. \cite{Arvanitaki:2009fg} that superradiant instabilties of rotating BHs can source GWs. They showed the remarkable ability of rotating BHs in enabling detection of particles that have been proposed with different theoretical motivations, but are elusive. Of particular interest are axion-like particles that are considered strong candidates for dark matter \cite{Arvanitaki:2009fg, Marsh:2015xka, Essig:2013lka, Hui:2016ltb}~. The first detection of GWs \cite{PhysRevLett.116.061102} by the Laser Interferometer Gravitational wave Observatory \cite{2015} (LIGO) has unsurprisingly boosted interest in understanding these sources. The purpose of this review is to give an overview of superradiant instabilities as sources of GWs and to assess the ability of current and future GW detectors in informing us about dark matter candidates.

In section 2 we will briefly outline the dynamics of these systems and how they are modelled as GW sources. Section 3 discusses the detectability of these sources, search strategies and search efforts by Advanced LIGO~\cite{2015} and Advanced Virgo~\cite{advvirgo} network of detectors (LV). Finally, in section 4 we discuss the impact of these clouds on other GW sources, particularly on standard GW sources - binary BH (BBH) coalescence.

\section{Superradiant Instabilities as sources of GWs}

A superradiant instability can develop for any bosonic field, whether it is a spin zero (scalar field ) or a spin non-zero field like dark photons (spin one) or a massive graviton (spin two). The field evolution equations on a Kerr background spacetime are much harder to solve for fields of nonzero spin in the relativistic regime. Although the instability grows much faster for a vector field than a scalar field, qualitatively it has the same features irrespective of the spin of the field. Motivated by this, we shall discuss the key features of the instability in the context of a scalar field here for simplicity. We encourage readers to familiarize themselves with the challenges of solving the equations and latest results for particles with non-zero spin \cite{Pani:2012bp, Baryakhtar:2017ngi, Dolan:2018dqv, PhysRevLett.120.231103, Britospin2, PhysRevLett.124.211101}~. 

The evolution of a massive scalar field, of mass $m_s$, on a generic metric is described by the Klein-Gordon equation
\begin{equation}\label{KGeq}
\frac{1}{\sqrt{-g}}\frac{\partial}{\partial x^\mu}\bigg(g^{\mu\nu}\sqrt{-g}\frac{\partial\Psi}{\partial x^\nu}\bigg) - \mu^2\Psi = 0,
\end{equation}
where $\mu$ is defined by $m_s = \mu\hbar$ . The product $\mu M$, $M$ being the mass of the BH, determines if the cloud is relativistic or not. For $\mu M \ll 1$ the cloud is non-relativistic which enables solving the radial part of the Klein-Gordon equation analytically as the equation now resembles a Schrodinger-like equation. This is no longer true in the relativistic regime and numerical calculations need to be employed. It is noteworthy that the cloud can accumulate at most 10 percent of the BH mass \cite{PhysRevLett.119.261101} and is spread out over a large volume. This allows us to treat the cloud as a perturbation to the BH spacetime, produced by the scalar field evolving on a Kerr background spacetime \cite{Brito:2014wla, Yoshino:2013ofa}~. These expectations were recently validated by nonlinear numerical evolutions for a spin-1 boson \cite{East:2017ovw}~. 

The scalar field instability has been studied quite extensively because of the relative ease in solving the Klein-Gordon equation on the Kerr metric. The radial and the polar components of the equation separate and give rise to a set of coupled differential equations that can be solved using numerical techniques \cite{Leaver:1985ax}~. Writing the scalar field as 
\begin{equation}\label{scalarsol}
\Psi = \int d\omega e^{-i\omega t+i m \phi}{}_0S_{\ell m \omega}(\theta)\psi_{\ell m\omega}(r)
\end{equation}
where ${}_0S_{\ell m \omega}e^{i m \phi}$ are the scalar spheroidal harmonics.
The azimuthal index, $m$, marks the energy level occupied by the bosons and $\omega = \omega_R + i\omega_I$ is the complex frequency of the bosonic field. Note that the real part of the frequency $\omega_R$ gives the frequency of oscillation of the field while the imaginary part $\omega_I$ determines the growth rate of the field. For superradiance to occur, $\omega_I$ must be positive, which translates into a constraint on $\omega_R$ and $\Omega_H$

\begin{equation}
\omega_R < m\Omega_H.
\end{equation}
To solve for the geometry of a BH spacetime sourced by the energy density of a scalar field, as obtained from the solution of the Klein-Gordon equation, one needs to solve the Einstein's equation.
%

Recall that we have already established that the scalar cloud is a perturbation to the background Kerr spacetime, so we essentially solve for the perturbation and calculate the GW flux from these systems. We shall not delve into how the Teukolsky formalism is used in this regard to avoid reproduction of an already well reviewed literature \cite{Teukolsky:1973, Sasaki:2003xr, Mino:1997bx}~. Instead, we shall focus on more recent works geared towards contextualising the GW signal to detectors. 

The timescale over which the instability grows and the timescale over which GW radiation takes place are well separated, this has been verified for a vector field by numerical simulations. This allows for an adiabatic treatment of the system. The key equations that govern the evolution of the instability are \cite{Brito:2014wla}
\begin{eqnarray}\label{cloudevol}
\dot{M} &=& -\dot{E_S}\nonumber \\
\dot{M} + \dot{M_S} &=& - \dot{E}\nonumber \\
\dot{J} &=& - m\dot{E_S}/\omega_R \nonumber\\
\dot{J}+\dot{J_S}&=&-m\dot{E}/\omega_R
\end{eqnarray}
where $E_S$ is the energy flux extracted from the BH, $M_S$ and $J_S$ are the mass and angular momentum of the cloud, respectively. An adiabatic evolution means that the second equation of Eq. \ref{cloudevol} becomes
\begin{equation}
\dot{M_S} = -\frac{dE}{dt}
\end{equation}
where $E$ is the GW energy flux, as the emission begins only after saturation of the  superradiance condition, i.e., once the BH has spun down enough such that $\omega_R = m\Omega_H$ and superradiance shuts off. A full non-linear evolution would typically have to account for the sub-dominant GW absorption at the horizon, GW emission before saturation and back-reaction of the cloud on the metric. Several efforts have been made and are also currently underway to produce numerical simulations \cite{Zilhao:2015tya, PhysRevD.89.104032, PhysRevLett.116.141101, PhysRevLett.116.141102, East, PhysRevLett.121.131104} that can follow the instability, but they are computationally expensive and even prohibitive in case of a scalar field. The timescale over which the instability develops for a vector field is more tractable than that of a scalar field, as growth rate is much faster for the former.

To assess detectability of these systems, the GW flux has to be translated to signal strength at detectors, taking into account the typical geometry of the detector. The waveform obtained by \refcite{Brito:2017zvb} was reinterpreted in a way that is more compatible to data analysis by \refcite{Isi:2018pzk}~. The signals emitted by these sources are long-lived and monochromatic, emitting at twice the frequency of the scalar field with a small positive drift in the frequency.
The GW strain at a detector can be written as 
\begin{align}
h = h_+F_+ + h_\times F_\times
\end{align}
where $F_{+/\times}$ are functions that encode the response of the detector to a signal, depending on the direction of arrival. These pattern functions are sensitive to the GW polarization; to be specific, two co-located detectors that are sensitive to the exact same polarization of a GW would have the same pattern functions. The GW polarization amplitudes, $h_+$ and $h_\times$, can be expanded on a spin-weighted spheroidal harmonic basis \cite{Brito:2017zvb}~,
\begin{equation}\label{charstrain}
h_{+/\times} = \sum_\ell h_0^\ell(r)\big({}_{-2}S_{\ell \tilde m \tilde\omega}(\theta)\pm (-1)^\ell{}_{-2}S_{\ell -\tilde m -\tilde\omega}(\theta)\big)\cos[\tilde \omega (r^*-t)+\phi_0+\tilde m\varphi].
\end{equation}
The radial function $h_0^\ell$ encodes the amount of energy in each mode and can be written as a characteristic strain amplitude, such that the dependence on the mass of BH and boson are explicit. Rewriting the radial function this way helps to isolate a dimensionless measure of energy in each mode, that is a function of BH spin and $\mu M$ alone. Note that $\tilde m = 2m$ and $\tilde \omega = 2\omega_R$. A factor of $(-1)^\ell$ before the second spheroidal harmonic in Eq.~(\ref{charstrain})~was missed in both \refcite{Brito:2017zvb} and \refcite{Isi:2018pzk}~, but later addressed in the Erratum of \refcite{Isi:2018pzk}~.
\begin{figure}[h]
\centerline{\includegraphics[width=3.2in]{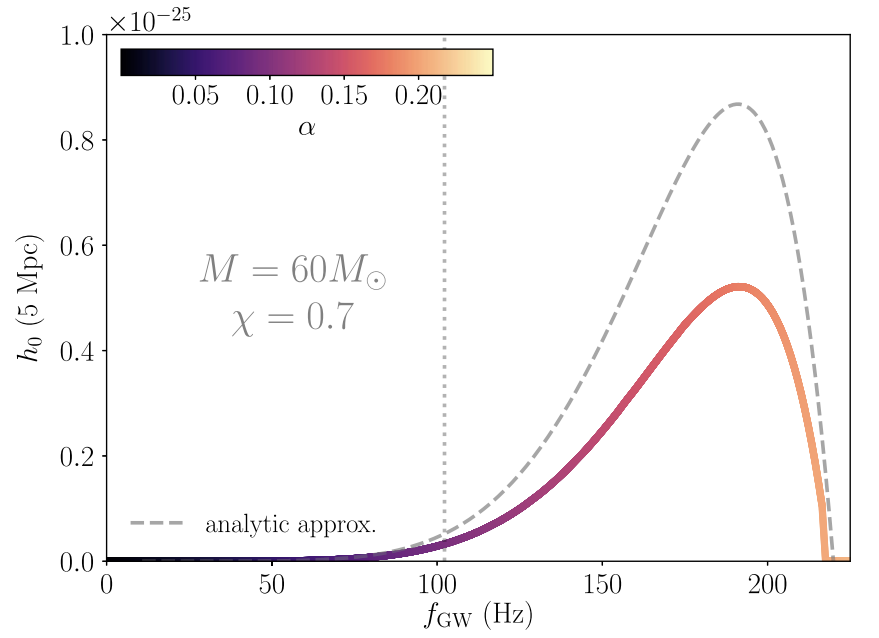}}
\vspace*{8pt}
\caption{Strain amplitude vs GW frequency for a fixed BH mass and spin, assuming that the source is at a distance of $5$ Mpc. The frequencies on the \textit{x-axis} correspond to different boson masses. The colored line shows the numerical estimates obtained from perturbative calculations of the characteristic strain amplitude and the dashed curve shows the analytical estimate, valid for $\mu M\ll1$. The color of the curve denotes the value of  $\alpha (= \mu M)$.~\textit{\textbf{Image courtesy:}}~Isi et al.\cite{Isi:2018pzk}~\protect\label{fig1}}
\end{figure}
\begin{figure}[h]
\centerline{\includegraphics[width=3.3in]{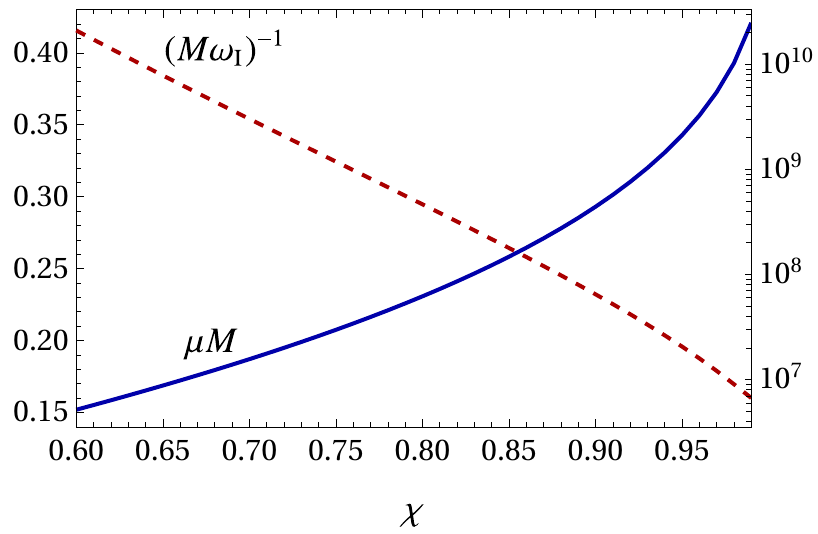}}
\caption{ Left \textit{y-axis} shows the $\mu M$ corresponding to the fastest growing instability (blue solid line) and right \textit{y-axis} shows the instability timescale (red dotted line) in natural units as a function of dimensionless BH spin $\chi$.~\textit{\textbf{Image courtesy:}}~Ghosh et al.\cite{Ghosh:2018gaw}~\protect\label{fig2}}
\end{figure}
The quantity $\mu M$ is critical in determining the intrinsic properties of the source, specifically the strength of the signal and timescale over which the instability develops. These features are well depicted by Figures~\ref{fig1} and~\ref{fig2} respectively. It is clear from these plots that the characteristic strain amplitude is not a monotonically increasing function of $\mu M$ i.e., superradiance shuts off  beyond an optimal value of $\mu M$, for a fixed BH mass. This arises from the interplay of the two constraints -- the superradiance saturation condition and the limit on BH spin. The instability timescale in Fig. \ref{fig2}~is BH-mass independent and can be used to estimate the instability timescale for the fastest growing mode, corresponding to a given spin at any BH mass. This serves as a useful guide in determining the potential of detecting the superradiant instability of a BH produced in a merger.

All the sources that are loud enough i.e., the signal amplitude is above the noise floor of a detector, constitute a detectable population. To determine rates of these events one has to use population models of relevant astrophysical sources, therefore the reliability of these results are always contingent upon the accuracy of the models used. The instabilities that are not loud enough to be detected may still be informative. All of the unresolved sources would produce a stochastic background that can be calculated as the ratio of the energy density of GWs from these sources per logarithmic frequency interval at the signal frequency as measured at present, to the critical density of the Universe \cite{Phinney:2001di,Brito:2017wnc}~,

\begin{equation}
\Omega_{GW} = \frac{1}{\rho_c}\int \frac{dE_{GW}}{d{\textrm{ln}f}} N(z) dz.
\end{equation}
The integration is over only the unresolved sources and $N(z)$ dz is the number of events in a unit comoving volume within a redshift interval of $z$ and $z+dz$.

These sources can be probed by several different search methods. Given that superradiant instabilities are ubiquitous to all Kerr BHs for a massive bosonic field satisfying the superradiant condition, the rate of these events is quite promising for detection \cite{Brito:2017zvb}~. So an \textit{all-sky} search, where a broad range of search  parameters can be probed is an obvious choice. Typical choice of search parameters for this kind of a search are the source location, frequency and drift in frequency of the signal. A \textit{directed} search can be conducted in case of an instability of a BH whose sky location are known either by electromagnetic observations or from GW observation of mergers. As noted above, all unresolved sources would generate a stochastic background that may itself be detectable \cite{Brito:2017wnc}~, therefore a search devised specifically for this signal is also well motivated. A detection from any of these searches will have to be followed up with inputs from particle physics in order to confirm the nature of the boson; non-detection on the other hand helps us constrain the plausible mass of the boson. All of these search strategies have been implemented on LV data and that has helped in setting important constraints on the boson mass. We discuss the details of the results in the following section.

\section{Detectability}

The prospect of detecting these sources by GW detectors was first discussed by Arvanitaki et al. \cite{Arvanitaki:2014wva}~. They explored these systems in a non-relativistic limit on a flat background spacetime, as a conservative estimate. This was followed up by more accurate numerical estimates calculated on a Kerr background, and the results for both the resolved and unresolved sources were found to be promising, even for existing ground-based detectors. Furthermore, an indirect method of detecting signatures of bosonic clouds is by looking at the spin distribution of isolated BHs. As the cloud grows by extracting rotational energy of the BHs, a universe with a massive boson could have a different spin distribution in the BH population than one without a boson. In this section we will discuss these three different detection possibilities, the state-of-the-art in search methodologies and current results.

\subsection{Stochastic background}

The strongest contender among the signatures discussed above is perhaps the self-confusion noise of these sources. The strength of signal for these bosonic clouds are modulated by the mass of cloud and the distance they are at from a GW detector. Numerical estimates suggest that the mass of the cloud is typically $\mathcal{O}(1)$ percent of the BH mass, with the most massive ones being $\sim10\%$. Therefore, the dependence of the signal strength on mass of the cloud is milder compared to the distance to the source. This means there will be a lot more unresolved sources than resolved ones, and all of these unresolved sources will contribute to a stochastic background.

This stochastic background noise was calculated for the first time in 
 \cite{Brito:2017wnc} and it was found to be really strong. In fact, based on the isolated BH populations considered in \refcite{Brito:2017wnc}~, the background was found to be loud enough (e.g., $\Omega_{GW}\sim10^{-6}$ for a boson of mass $\sim3.16\times 10^{-13}$eV) for detection in LIGO data from the first observing run (O1). 

This observation was followed up by \cite{Tsukada:2018mbp} in significant detail. Not only did they consider the background due to all unresolvable isolated BH-boson condensate but also the background due to clouds formed around remnants of BBH mergers. For isolated BHs they consider a few different spin distributions that can broadly be classified in a similar fashion as \cite{Brito:2017wnc} viz. optimistic and less optimistic scenario. For the merger remnant background, they assume that all of the remnants have a spin of 0.7, which is certainly a bit conservative  considering that latest GW catalog~\cite{PhysRevX.9.031040, PhysRevX.11.021053} reports multiple detections with higher remnant spins. As is evident from Fig. 2 of \cite{Tsukada:2018mbp}~, even a slight enhancement in the background from merger remnants due to a boson of mass $10^{-13}$ eV can push it into the detectable regime. With similar considerations for population models as \cite{Brito:2017wnc}~, they report a non-detection of the background in O1 data thereby excluding the mass range $[2-3.8]\times10^{-13}$eV.  An interesting possibility addressed by this work was to explore the ability of distinguishing the background from bosonic clouds and distant compact binary coalescence (CBC). They consider two models for the background, CBC-only and background from CBC and bosnic clouds, and try to gather support over the parameter space for one model over the other by looking at the log Baye's factor. Based on their analysis, they conclude that for a population with spins $[0.6-0.8]$ it may be possible to distinguish between a CBC-only and a mixed background at design sensitivity of LIGO-Virgo-KAGRA~\cite{KAGRA:2018plz} network of detectors (LVK). 

A similar analysis was done for vector bosons \cite{PhysRevD.101.024019, PhysRevD.103.083005} using data from the first two observing runs of LV. No signal was found, thereby disfavoring masses lying in $[0.8 - 6.0]\times 10^{-13}$ eV for a vector boson. However, such exclusions are always to be interpreted with caution, bearing in mind the uncertainties introduced by population models (see Fig. 2 on \refcite{Brito:2017wnc}) and other considerations. 

Some of these caveats were explored by \refcite{Yuan:2021ebu}~. Apart from the impact of population models on detectability by third generation of ground-based detectors, they also looked at the contribution from higher modes of these systems. As mentioned earlier, the BH-boson cloud system can be thought of as a gravitational atom, with levels marked by the indices $\ell$ and $m$. The $\ell=m=1$ level leads to the fastest growing instability, but in principle, all other levels can undergo a superradiant instability. As long as these higher modes undergo instability within the age of the Universe, they can contribute to the total GW flux emitted by the system. They estimate the total GW flux due to all $m\geq1$ modes under two reasonable, simplifying assumptions: 1) At any given time, only one mode undergoes the instability. It has been shown by \refcite{Dolan} that the timescales of the growth of the instability of different levels are well separated 2) For each $\ell$, the $\ell = m$ is the fastest growing one and therefore the primary contributor to the background.

\begin{figure}[h]
\centerline{\includegraphics[width=4.95in]{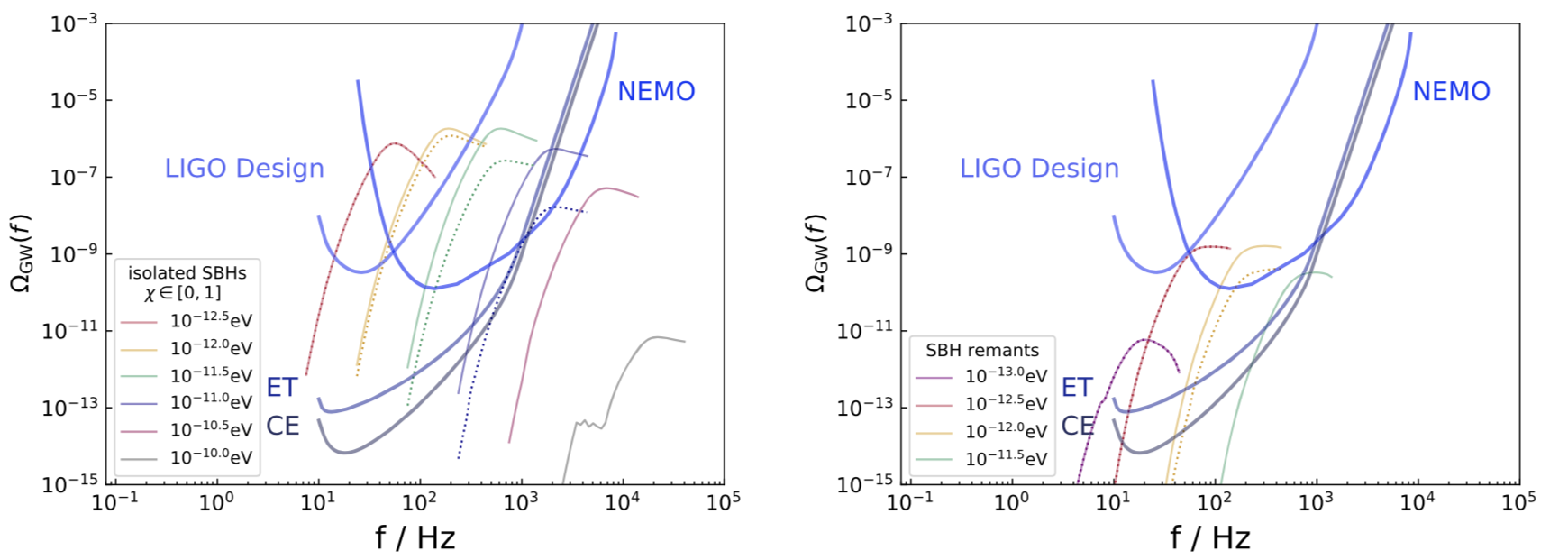}}
\vspace*{8pt}
\caption{\textit{Left panel}: Total background from clouds formed around isolated BHs with uniformly distributed spins $\in[0,1]$ and merger remnants. \textit{Right panel}: Background from instabilities of BHs formed through binary mergers only. Dotted lines show the contribution from the fastest growing mode, $m = 1$, and solid lines show cumulative contribution from all relevant $m \geq$ 1. Power-law integrated sensitivity curves for various detectors are shown and indicated with their names.  ~\textit{\textbf{Image courtesy:}}~Yuan et al.\cite{Yuan:2021ebu}~\protect\label{fig3}}
\end{figure}

While the impact of higher modes is negligible for boson masses that can be probed by LVK, left panel of Fig. \ref{fig3} makes a compelling case  for adding the contribution from higher modes to the background for  bosons with $m_s>10^{-12}$eV. In particular, for bosons with masses $\gtrsim10^{-11}$eV, the background is found to be almost entirely due to higher modes. This makes the higher modes particularly interesting for third generation of ground-based detectors like Einstein Telescope \cite{ET} (ET), Cosmic Explorer \cite{LIGOScientific:2016wof} (CE) and Neutron Star Extreme Matter Observatory \cite{Ackley:2020atn} (NEMO). 

Yuan et al. simultaneously highlights the importance of uncertainties in population models and of adding the contribution from merger remnants with the right panel on Fig. \ref{fig3}. The mass range ruled out by non-detection of the background in LV data relies upon a uniform distribution of the isolated BH spins. For a very pessimistic scenario where all isolated BHs are presumed to be born with really low spins, such that the vast majority of them do not undergo superradiant instability in their lifetime, the main contribution to the background comes from remnants of BBH mergers. They conclude that this contribution might be marginally detectable by LIGO at design sensitivity and would be picked up quite easily by 3G detectors. 

\subsection{Resolvable events}

\subsubsection{Event rates}

For long-lasting signals that are shorter than the observation time of a detector, event rates can be calculated using \cite{Brito:2017zvb}~,

\begin{equation}
N = T_{\textrm{obs}}\int_{\rho>\rho_{\textrm{th}}}\frac{1}{1+z}\frac{d^2 \dot{n}}{dMd\chi}\frac{d V_c}{dz}dz~dM~d\chi .
\end{equation}

Here $\rho_{\textrm{th}}$ is the signal-to-noise ratio threshold for claiming detection. This is essentially obtained by integrating the BH comoving-volume number density (see \refcite{Brito:2017zvb} for details of population model used) over a comoving volume for the resolvable sources only. 

Rates for both bosonic clouds around massive BHs and stellar mass BHs were considered. Since the mass of the boson determines the frequency of the signal, planned space-based detectors like Laser Interferometer Space Antenna (LISA) and ground-based detectors are sensitive to different ranges of mass of boson. Fig. 6 on \refcite{Brito:2017zvb} shows the expected event rates for a range of boson masses, spanning both the LIGO and LISA range. Note that the stochastic background from unresolvable sources contribute to noise budget for the detectable sources. The impact of this additional noise on event rates is shown by the thin lines on this figure. The peak of the rates obviously occurs at the frequency of peak sensitivity of the detector. For long lasting signals like this, a semi-coherent search method is often preferred. Tables I and II on \refcite{Brito:2017zvb} summarizes the expected range of event rates for optimistic to pessimistic BH population scenarios and for the more computationally efficient, but pessimistic semi-coherent search method. It is clear from these numbers that even under the most conservative assumptions at least $\mathcal{O}(1)$ events are expected at LISA for a boson mass of $10^{-17}$ eV and at LIGO for a boson mass in $[10^{-12},10^{-12.5}]$ eV.

\subsubsection{All-sky searches}

The data analysis techniques used for these signals are similar to other long-lived, monochromatic signals like GWs emitted by spinning neutron stars with asymmetries. Using the \textit{all-sky} search results \cite{LIGOScientific:2019yhl} for periodic signals on real data from the second observing run of LV (O2) no candidate of significance was found \cite{Palomba:2019vxe}~. However, even a non-detection is useful in setting upper limits on the strength of the signals, that translates into constraints on the boson mass. Based on these results a boson mass range of  $[1.1 \times 10^{-13} - 4 \times 10^{-13}]$ eV was excluded for an optimistic BH spin distribution, and $[1.2 \times 10^{-13} - 1.8 \times 10^{-13}]$ eV was excluded for a less optimistic spin distribution (see \refcite{Palomba:2019vxe} for details). The minimum amplitude sensitivity in the first half of the third observing run of LV (O3) was $1.6$ times lower than O2 \cite{PhysRevD.103.064017}~, which would potentially improve the existing constraints. 

\subsubsection{Directed searches}

In case the location of the source is known, it is possible to conduct a directed search in the data. This can be implemented either by following up the remnant from a binary merger detected through GWs or by focussing on BHs known from electromagnetic observations, such as BHs in X-ray binaries \cite{PhysRevD.95.043001}~. Following up a merger remnant can be helpful only if the instability grows fast enough for it to start emitting GWs within the observation time of the detector. Therefore, it is possible to \textit{a priori} carve out regions of the BH mass-boson mass parameter space that can be probed by a detector, for a specific choice of instability growth time. For instance, Fig. 4 on \refcite{Ghosh:2018gaw} shows the BH-boson masses that would produce a detectable signal at the planned future detector CE, provided the instability grows within thirty days of the merger. For highly spinning remnants a larger part of the parameter space can be probed as the signal strength is modulated by the initial spin of the BH. This kind of a plot can serve as a guide to determine the minimum observation time required for detecting the instability for a BH of a particular mass. Furthermore, in case of non-detection this plot can be used to rule out a range of boson masses that lies within the contours.  

For a galactic population, it is pertinent to assume that multiple detectable signals will overlap together to form an \textit{ensemble} signal~\cite{Zhu:2020tht}~. Since all the sources will radiate at nearly the same frequency, such a signal may be harder to detect by existing continuous wave searches, and an all sky narrowband radiometry on folded data may be helpful \cite{PhysRevD.91.124012,PhysRevD.98.064018}~.  A rigorous characterization of such a signal by simulating populations of isolated BHs within the galaxy was done by \refcite{Zhu:2020tht}~, taking into account the Doppler shift of the sources and making appropriate considerations for the signal strength as a function of the GW radiation timescale. Another interesting search methodology that can be implemented for third generation of detectors is to conduct multiband searches for superradiant instabilities \cite{Ng:2020jqd}~. The idea is to use the sky location obtained from the inspiral signal detected by LISA and then carry out a directed search for the continuous GWs emitted by a bosonic cloud around one or both of the components, in CE data. 

A hidden Markov Model tracking method, originally developed for rapidly spinning neutron stars, can be used for directed searches of these sources \cite{Isi:2018pzk}~. This method is more versatile than a semi-coherent or coherent search method as it allows for small deviations from a stringent signal model (see \refcite{Isi:2018pzk} and references therein for the details of the search algorithm). Fig. 9 on \refcite{Isi:2018pzk} demonstrate the robustness of the algorithm to uncertainties in the signal model. They also show the farthest distance (detection horizon) to which an optimal boson cloud (fastest growing level) can be detected by present and future ground-based detectors in Fig. 12. The detection horizon for LV, even at design sensitivity, is about $20$ Mpc which is much smaller than the nearest merger seen by LV to date. Therefore, following up remnants of mergers seen by LV is perhaps not conducive to constraining the mass of the boson. However, third generation of ground-based detectors, like the CE will have excellent potential of detecting boson clouds by following up mergers, potentially picking up signals $10^4$Mpc away while probing a sizeable range in BH-boson masses.

One of the challenges in case of a directed search is the uncertainty in sky location. Mergers detected by only two detectors have significantly poorer localization than ones detected by three detectors. BHs in X-ray binaries have an advantage in this aspect, since they are quite well located on the sky. However, due to large uncertainties in their spin and also their history, constraints on boson masses derived from these BHs are subject to important caveats. A directed search on Cygnus X-1, assuming an age of $5\times 10^6$ years and near-extremal birth spin found no signals consistent to a superradiant instability \cite{PhysRevD.101.063020}~. The analysis is done in a BH spin-agnostic way, which allows them to conservatively constrain the disfavored boson mass range to $[6.4,8.0]\times 10^{-13}$ eV. It is worth reiterating that these constraints rely heavily upon the accuracy of the measurements of the source properties, as demonstrated by Fig. 3 of  \refcite{PhysRevD.101.063020}~.

\subsection{Spin-distribution of BH binaries}

So far we have discussed direct detection of GW signals from BH-boson cloud systems. However, since a superradiant instability grows by extracting rotational energy from a BH, a boson would leave its imprints on the spin distribution of astrophysical BHs. This should show up as holes on the BH mass-spin plane i.e., the Regge plane \cite{arvanitaki83}~, and if a BH is discovered, by GW or electromagnetic observations in these excluded regions then that disfavors the existence of a boson of that mass. Of course this holds true only if the BH mass and spin can be determined with great precision. Based on spin measurements from electromagnetic observations, roughly the range $[6\times 10^{-13},10^{-11}]$ eV is disfavored \cite{ Arvanitaki:2014wva, PhysRevD.98.083006, Cardoso_2018} but, again, it is important to note the uncertainty in the spins. Other scenarios, using spin estimates from the first imaging \cite{EventHorizonTelescope:2019dse} of a BH by the Event Horizon Telescope \cite{Davoudiasl:2019nlo}~ and using spin estimates from a X-ray binaries and GW observations together \cite{Stott:2020gjj, Ghosh:2021zuf} to constrain the mass of ultralight bosons have also been explored.

While the spin of the remnant may be well determined for LV sources, the spins of the components in a BBH merger are typically quite poorly constrained. For LISA sources, since the spins of the BHs will be determined to great accuracy a Bayesian model selection can be performed to determine if the BH population favors the existence of a boson \cite{Brito:2017zvb}~. An interesting approach for LV sources would be to use a hierarchical Bayesian inference technique to glean information about the boson mass and the distribution of BH spins at formation \cite{PhysRevD.103.063010}~. This is essentially done by considering two models, one with a boson that couples with stellar mass BHs in binaries and spins them down before they merge, and another where the BBHs are born with lower spins at formation itself. With enough BBH merger events it is possible to identify which model is preferred by the data, by looking at the joint posterior distribution of the boson mass and spin distribution of the component spins at formation. This was indeed done by \refcite{PhysRevLett.126.151102} using all of the events in the GW catalogs \cite{PhysRevX.9.031040, PhysRevX.11.021053} released by the LVK Collaboration. Their findings suggest that a boson in the mass range $[1.3-2.7]\times10^{-13}$ eV is strongly disfavored by two of the 45 events detected by LVK until the first half of O3 (see Fig.2 on \refcite{PhysRevLett.126.151102}).  It is important to note that this is a statistical method and therefore the accuracy of the constraints will depend on the number of events used in determining them. 

\section{Impact on other sources}

\subsection{Stellar mass black hole binaries}
In the previous section we discussed the scenario of bosonic clouds spinning down the BHs in stellar BH binaries. In reality BH binaries dressed in a cloud will undergo a lot more interesting dynamics than just a simplistic spin down \cite{PhysRevD.96.083017, PhysRevD.99.064018}~. As mentioned before, the fastest growing instability corresponds to $\ell=m=1$; on the other hand $m=0,-1$ correspond to decaying modes. It so happens that the gravitational perturbation introduced by the companion BH may lead to mixing between the growing and decaying modes \cite{PhysRevD.99.044001,PhysRevD.101.083019}~. Such a phenomenon would leave its imprints on both the GW signal from the cloud as well as the binary inspiral. Both analytical estimates for a non-relativisitc cloud of circular binaries \cite{PhysRevD.99.044001} and perturbation theory calculations for a relativistic cloud of non-circular binaries \cite{PhysRevD.99.104039, PhysRevD.101.043020} highlight the impact of these effects on cloud depletion. 

Given that both of these treatments are carried out under multiple simplifying assumptions, fully numerical simulations are essential in understanding the actual dynamics of these systems. This is quite a challenging problem to be handled by existing numerical relativity (NR) codes as they evolve only the late inspiral to merger portion of a BBH coalescence. The analytical/numerical estimates suggest that cloud may be significantly depleted quite early on therefore one needs to know the cloud profile very well at the separation where the NR simulations start. Note that the quantity $\mu M$ determines the size of the cloud, therefore a non-relativistic cloud ($\mu M\ll1$) would be quite far away from the individual BHs. So, a full numerical treatment from early inspiral is absolutely necessary to estimate the impact on the inspiral-merger-ringdown (IMR) waveform of a BBH merger for any realistic data analysis. On the other hand for $\mu M\lesssim1$ would be much more closely bound to the individual BHs, and therefore it is important to consider the back-reaction of the energy density of the cloud on the metric particularly in the late inspiral-merger region. Several efforts are currently underway in trying to understand how the presence of a cloud around one of the components in a binary \cite{PhysRevD.101.043020, PhysRevD.101.064054, Traykova:2021dua, PhysRevD.103.044032, Annulli:2020lyc, Baumann:2019ztm} or a cloud enveloping a binary \cite{PhysRevD.103.024020} would impact the structure of the cloud and the evolution of the binary itself.

\subsection{Massive black hole binaries}

Apart from the possibility of detecting bosonic clouds of isolated massive BHs at LISA, there are other ways in which instabilities of massive BHs may be interesting. Obviously, the mixing between growing and decaying modes would be relevant for massive BH binaries as well. But an even more fascinating scenario may be offered by another kind of LISA source --  an extreme mass ratio inspiral (EMRI) of a small compact object around a massive BH. It may be possible to confirm or rule out the existence of a boson of a particular mass by just a single EMRI observation  at LISA \cite{Hannuksela:2018izj}~. Masses and spins of massive BHs will be determined very precisely by LISA through GW observations, which gives an estimate of the boson mass that can couple efficiently with the massive BH via the superradiant condition. On the other hand, since the smaller object will orbit the massive BH dressed in a cloud for thousands of cycles, the GW signal will encode information about the gravitational potential due to the cloud, providing an independent measurement of the boson mass. If the two estimates are contentious then that rules out the boson mass, and if they agree, that simultaneously confirms the existence of a boson and gives an accurate estimate of its mass.

\section{Summary and outlook}

The idea of superradiant instabilities of rotating BHs has been around for nearly half a century. However, the inability to probe these systems electromagnetically rendered them beyond the scope of observation. The identification of these systems as potential GW sources has been pivotal in generating interest in this field of work, and now that we are in the GW detection era, looking for these sources is a realistic goal. 

In this review we have restricted ourselves to discussing a bosonic field without non-linear self-interactions. It is important to note that adding self-interaction may impact GW emission adversely for large enough interactions \cite{PhysRevD.103.095019}~, and a large fraction of astrophysical BHs may never reach the interesting final stage of a \textit{bosenova} \cite{Yoshino:2012kn}~. Furthermore, if a coupling between the scalar field and photons is considered, that could also limit the growth of the cloud thereby reducing GW emission \cite{PhysRevLett.122.081101}. However, this may have an interesting consequence with the source emitting bursts of light through laser-like emission \cite{PhysRevLett.120.231102, PhysRevD.98.103012}~.

We have discussed the various direct and indirect methods of looking for bosonic clouds of stellar BHs and massive BHs. There has been remarkable progress in terms of developing data analysis pipelines for these sources in the recent past. A lot of these techniques will be useful for future, more sensitive ground-based and space-based detectors. For the stellar BHs we have reviewed the results of the different searches that have been conducted on LV data, and highlighted the caveats for each of the constraints. Statistical studies of BH spin distributions complemented by Regge plane holes have been demonstrably used to constrain boson mass and these bounds will only improve in the subsequent observing runs of LVK.

An important and emerging line of research in this regard is the study of the modulation of the IMR signal due to the presence of a cloud. Fully non-linear numerical simulations would help immensely in gauging the need of introducing effects of the cloud on a binary in existing IMR waveform models. 

Observing a signal from the superradiant instability of a rotating BH would be exciting on its own merit. But the possibility of detecting particles outside the Standard Model, particularly ones that could potentially be constituents of dark matter, makes it even more appealing. With more advanced detectors in the horizon, the future of this field is exciting and full of potential.

\section*{Acknowledgments}

We would like to thank Mark Hannam, Richard Brito and Shaon Ghosh for useful comments. We thank our colleagues for very kindly letting us include figures from their work in this review. S. G is supported in part by Science and Technology Facilities Council (STFC) grant ST/V00154X/1 and by European Research Council (ERC) Consolidator Grant 647839. 

%

%
%
%
%

\bibliographystyle{ws-mpla}
\bibliography{sghosh-mpla}
\end{document}